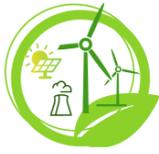
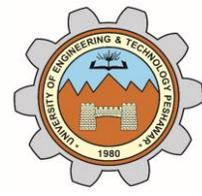

# Impact of Thermal Variability on SOC Estimation Algorithms

Wasiue Ahmed [1, *], Mokhi Maan Siddiqui[1], Dr. Faheemullah Shaikh[1]

[1]*Departmnet of Electrical Engineering, Mehran University of Engineering and Technology, Jamshoro 76062, Pakistan*

*\*Corresponding author*
***Email***: *wasiueabbasi@gmail.com*

**ABSTRACT**

While the efficiency of renewable energy components like inverters and PV panels is at an all-time high, there are still research gaps for batteries. Lithium-ion batteries have a lot of potential, but there are still some problems that need fixing, such as thermal management. Because of this, the battery management system accomplishes its goal. In order for a battery management system (BMS) to function properly, it must make accurate estimates of all relevant parameters, including state of health, state of charge, and temperature; however, for the purposes of this article, we will only discuss SOC. The goal of this article is to estimate the SOC of a lithium-ion battery at different temperatures. Comparing the Extended Kalam filter algorithm to coulomb counting at various temperatures concludes this exhaustive investigation. The graphene battery has the highest SOC when operated at the optimal temperature, as determined by extensive analysis and correlation between SOC and temperature is not linear.

## 1    INTRODUCTION

Energy storage and use are crucial during the switch to green energy. When we talk about energy storage, we immediately think of batteries. The battery is the main part that stores energy and needs to be reliable, effective, and healthy. These things can be guaranteed by the battery management system. Batteries need to be guarded against high voltage and dangerous operating situations. The system not only keeps an eye on and takes care of the battery modules, but it also sends out early warnings. So, the BMS makes sure the batteries work the way they are supposed to. An important part of battery management systems is figuring out how full a battery is. It aids in describing the actual amount of energy stored in the battery, i.e., State of Charge (Plett, 2004). Assessing the battery's SOC is just as important as knowing how much long it will last. There are several methods for estimating the battery's state of charge (SOC). According to (M. Mastali et al., 2013) the Equation (1), states that battery's state of charge is measured as the fraction of its nominal capacity Qn that is available at any given time, denoted by the capacity Q(t).

$$SOC(t) = Q(t)/Q_n \qquad (1)$$

As stated by (J. Meng et al. 2018), a proposed comprehensive study comparing various SOC estimation algorithms and their online implementation, the Kalman filter is the most promising method, considering calculation power and sacrificing efficiency. Comparative research using EKF and Sigma Point Kalman Filter was also conducted by (Jiahao Li, 2013), by establishing the initial SOC, the study discussed the tracking accuracy performance. In addition (Joaqun Klee Barillas et al., 2015), compared and analysed the cost of various algorithms, recognizing that as computational power increases the efficiency increase, so does the cost. Therefore, we must settle for some middle ground. Numerous other researchers have developed methods for accurately estimating SOC, such as (Shi Li et al., 2019), who proposed a comparative study between four algorithms and estimated the SOC by comparing the different load profiles, demonstrating that algorithm performance is dependent on the load patterns and is correlated in frequency spectrum. In the literature, the SOC of lithium-ion batteries has been estimated under a variety of conditions, but the scope of this article is to analyse SOC under various thermal conditions.

## 2   METHODOLGY

The SOC estimation algorithm is utilised in conjunction with other battery management techniques to prevent the battery from being over-discharged or over-charged and to prolong its life. Researchers have focused on the challenge of SOC estimation, leading to the development of numerous approaches. Classifying the methods is difficult because most solutions involve using multiple techniques at once and incorporating either heuristic or deterministic mathematical tools. It will be shown that the methods of coulomb counting (CC) and open circuit voltage (OCV) are utilised frequently by discussing both. Because individual methods can have their own shortcomings, combining them can result in a wide range of enhancements to both the initial and online SOC estimation. For instance, a robust extended Kalman filter algorithm (EKF) could be combined with the OCV method and the CC method as the secondary function. It becomes more complicated to categorise individual methods when they are combined in this way. In further sections, we will be discussing both the methods used in combined algorithms to estimate SOC at different thermal conditions.

### 2.1   Coulomb Counting Method

The use of CC for SOC estimation has become the standard. Because it is the most accurate method for short-term calculations, the CC method (ampere hour method) is defined in the Equation (2).

$$SOC(t) = SOC(t_o) + \frac{1}{C_n} \int_{t_o}^{t_o+t} I_{bat}(d\tau) * 100\% \qquad (2)$$

In the equation (2), $SOC(t_o)$ represents SOC at time $t_o$, $C_n$ represents specified capacity, and $I_{bat}$ represents the current. CC can be implemented easily, due to initial SOC there are error and challenges to consider. When measuring battery current, sometimes an error occurs in calibration and a proper current curve is not obtained. Accumulated errors are caused by noise, the wide range in sensor resolution, or rounding. Accumulated errors make equations less accurate over time, so they need algorithms to help them out. In actual practice, the initial SOC is unknown, and the initial SOC of a battery can only be found when the battery is in thermodynamic equilibrium. The charge and discharge current time integrals are evaluated to estimate SOC by the Coulomb method, and the initial SOC value is needed. Most of the time, it is unknown, and intentionally set to the wrong value. This method is dependent on the initial SOC value and cannot eliminate cumulative errors. If evaluated at the wrong SOC, we will get improper results. Despite its widespread use in recent years, the CC method is not typically used as a stand-alone technique for estimating SOC but rather in conjunction with other techniques. This paper employs CC along with EKF to estimate SOC.

### 2.2   2RC Electrical Circuit Model Based Estimation (2RC-ECM)

Most of researchers have used ECM models for SOC estimation, in many contexts, the second order resistor capacitor (2RC) ECM is used because of its ease of use and precision. For this purpose, we will be analysing the 2RC-ECM. Figure 1 depicts the 2RC-ECM. There is a voltage source (the OCV), an ohmic resistance (Ro), and a resistance and capacitance poles (RC). Both the activation polarization resistance and capacitance are represented by R1 and C1 in the resistor capacitor branches. Similarly, R2 and C2 stand in for the resistance and capacitance of the concentration polarization, respectively. The current through the load is denoted by I. The 2RC-ECM and Kirchhoff's law allow us to rewrite the battery model as a state space model as in (Rivera Barrera JP et al., 2017). The two voltages for the resistor capacitor forks are denoted by U1 and U2, defined in equations (3) & (4) respectively. Capacity of the battery, denoted by q. Equation (5) defines the SOC in terms of battery capacity, and Equation (6) depicts the relationship between open circuit voltage and state of charge.

$$U_1 = -\frac{1}{R_1 C_1} U_1 + \frac{I}{C_1} \tag{3}$$

$$U_2 = -\frac{1}{R_2 C_2} U_2 + \frac{I}{C_2} \tag{4}$$

$$SOC = -\frac{I}{q} \tag{5}$$

$$U = OCV(SOC) + U_1 + U_2 + IR_o \tag{6}$$

As seen in Figure 1, the battery ECM under consideration. It features a double pole RC circuit, compared to single and triple RC structures, which offers the optimal balance among inaccuracy and complexity also it is one of the most popular ECMs.

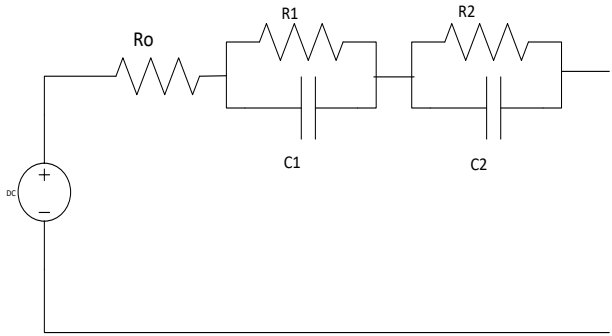

Figure 1: Second Order RC Model of Li-ion Battery

## 2.3 HPPC Test

The Hyper Pulse Power Characterization test has been conducted by Dr. Phillip on Turnigy Graphene 5000mAh 65C Lithium-ion Battery. The test was carried out at six different temperatures, but in this article only four temperatures have been selected for simplicity (Kollmeyer et al., 2020). The data is real time and is utilized in this article for SOC Estimation. Battery capacity is calculated using the static capacity test described above, and then 10% SOC steps are used for the HPPC test. The steps are taken from full SOC to no SOC, and then the test profile can resume at full SOC.

Included in each stage are:
- Wait an hour before continuing.
- Discharge, pulse duration: 1C for 10 seconds
- (Relaxation period) a 10-minute break.
- Regeneration Pulse: 1C or 0.75C for 10 seconds
- (Relaxation period) Take a 10-minute break.
- Discharge/charge the next step per the manufacturer's data sheet.

Each type of cell can be tested at each SOC, as well as at different temperatures and charge and discharge rates (C-rates).

## 2.4 The Extended Kalman Filter

The nonlinear system is linearized through the extended Kalman filtering (EKF) algorithm. It makes a prediction for the value of the following time step based on the current time step. For the best estimation, the state variables are continuously updated with data from the system's inputs and outputs. The battery estimation process calls for white noise with a Gaussian distribution, both in terms of the processing noise and the observation noise. This is a problem with every version of the Kalman filter. Here, it's simple to regulate the correlation between processing and observational noises. This method uses two step predication correction algorithms, as depicted in equation (7) & (8), where k denotes a discrete point in time, K is Kalman gain, P is covariance, Q is covariance of the process, and R is the covariance of the output as proposed by (F. Khanum et al., 2021).

$$\hat{x}_{k+1|k} = A\hat{x}_{k|k} + Bu_k \qquad (7)$$
$$P_{k+1|k} = AP_{k|k}A^T + Q_k \qquad (8)$$

The equation (9) calculates the Kalman Gain, and equation (10) updates the estimation with new measurement value, finally the equation (16) depicts the error covariance.

$$K_{k+1} = P_{k+1|k}C^T\left(CP_{k+1|k}C^T + R_{k+1}\right)^{-1} \qquad (9)$$

$$\hat{x}_{k+1|k} = \hat{x}_{k+1|k} + K_{k+1}(z_{k+1} - C\hat{x}_{k+1|k}) \qquad (10)$$

$$P_{k+1|k+1} = (1 - K_{k+1}C)P_{k+1|k} \qquad (11)$$

The algorithm has been modified to estimate SOC at different temperatures and associated errors in MATLAB.

## 3    SIMULATION & RESULTS

This article showed two different ways to estimate SOC accurately. Yet, the BMS is an area that needs further investigation. The main goal of this research is to make a better BMS that can accurately predict how long a Lithium-ion battery will last. But simulation is a good way to learn how dynamic systems, like Li-ion, act in different situations. For example, in our article we analysed the SOC of graphene batteries at four different temperatures using two different methods. In this article, the effect of temperature on SOC is looked at in depth. The foundation for this section is laid with the help of simulation and computational techniques discussed in the previous section. The Table 1 shows the comparative analysis of SOC at different temperatures using two different methods.

| Temperature | SOC by CC Method | SOC by EKF Method |
|---|---|---|
| 0 °C | 55.9785% | 52.3885% |
| 10 °C | 54.0007% | 52.8973% |
| 25 °C | 52.9031% | 55.0667% |
| 40 °C | 53.1736% | 54.9699% |

Table 1: Average SOC using CC & EKF methods at different temperatures

For more depth analysis, the graph is plotted by combing the SOC results of both methods, which is shown in Figure 2.

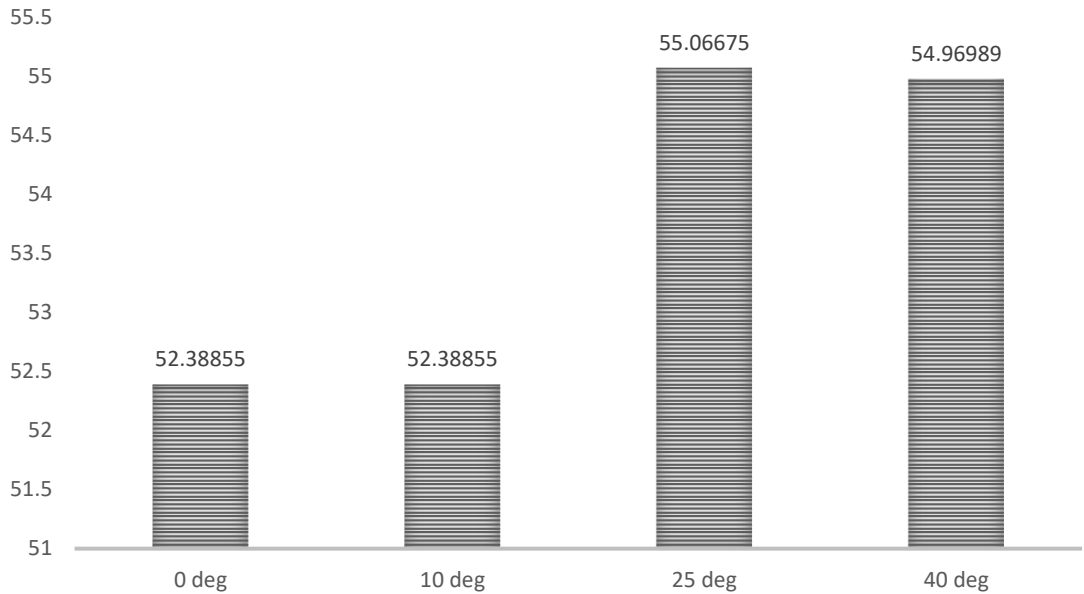
Figure 2: % SOC at Different Temperatures of Turnigy Graphene 5000mAh 65C Li-ion Battery

The result clearly shows that SOC rises as temperature rises. When the temperature reaches a certain point, the percentage of SOC will begin to drop. In addition to this, total thirty hours' time frame is implemented to observe the behavior of battery. Finally, to find the accuracy of methods used, the graph between average percentage error is plotted as visible in Figure 3, its evident that at near ambient temperature the error is relatively less.

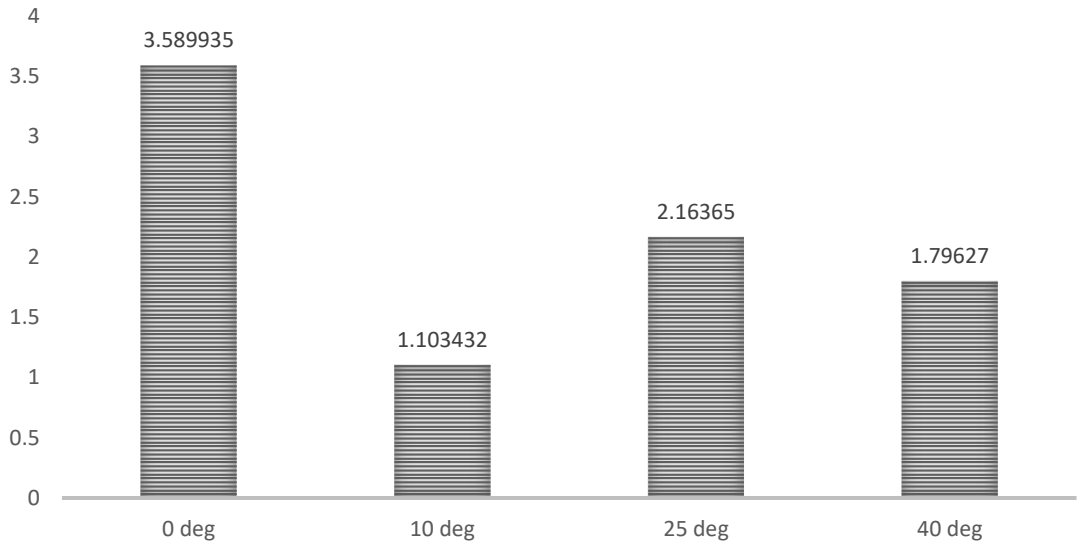
Figure 3: % Error in Estimated SOC at Different Temperatures

# 4	CONCLUSION

The necessity of this research is presented as a central idea. Following the execution of two distinct algorithms and the collection of their outputs, we arrived at the realization that the SOC is being estimated with an extremely small margin of error. Additionally, the nature of operation of lithium-ion batteries can be deduced from the results at various temperatures. The percentage of SOC begins to rise when the temperature does. When the temperature is raised even further, the SOC level begins to drop. This indicates that there is no direct correlation between state of charge and temperature, but it is best practice to keep batteries operating at a temperature that is close to 25 degrees Celsius. If someone wants to improve the accuracy of the results, this algorithm can be paired with other machine learning algorithms, which can further increase the efficiency as well as the accuracy of implementation. The implemented algorithm can be used while designing the BMS for a lithium-ion battery. Because the algorithm is straightforward but effective, it can easily be implemented on microcontrollers using any programming language.